\documentclass[amsmath,amssymb,aps,twocolumn,prl,longbibliography,superscriptaddress]{revtex4-2}

\usepackage{amssymb}
\usepackage{graphicx}
\usepackage{dcolumn}
\usepackage{bm}
\usepackage{amsmath}
\usepackage{subfigure}
\usepackage{color}
\usepackage[colorlinks=true,citecolor=blue]{hyperref}
\usepackage{multirow}
\usepackage{setspace}
\usepackage{comment}
\usepackage[normalem]{ulem}
\usepackage[utf8]{inputenc}
\usepackage[english]{babel}
\usepackage{xr}

\usepackage{soul}
\usepackage[bottom]{footmisc}
\hypersetup{allcolors=magenta}
\usepackage{xcolor}
\usepackage{hhline}
\usepackage{tabularx}

\begin{document}

\preprint{APS/123-QED}

\title{Arbitrary Instantaneous Bandwidth Microwave Receiver via  Scalable Rydberg Vapor Cell Array with Stark Comb} 

\author{Yuechun Jiao}
\affiliation{State Key Laboratory of Quantum Optics Technologies and Devices, Institute of Laser Spectroscopy, Shanxi University, Taiyuan 030006, China}
\affiliation{Collaborative Innovation Center of Extreme Optics, Shanxi University, Taiyuan 030006, China}
    
\author{Yuwen Yin}
\affiliation{State Key Laboratory of Quantum Optics Technologies and Devices, Institute of Laser Spectroscopy, Shanxi University, Taiyuan 030006, China}

\author{Yunhui He}
\affiliation{State Key Laboratory of Quantum Optics Technologies and Devices, Institute of Laser Spectroscopy, Shanxi University, Taiyuan 030006, China}
\affiliation{Collaborative Innovation Center of Extreme Optics, Shanxi University, Taiyuan 030006, China}
 
\author{Jinlian Hu}
\affiliation{State Key Laboratory of Quantum Optics Technologies and Devices, Institute of Laser Spectroscopy, Shanxi University, Taiyuan 030006, China}
 
\author{Cheng Lu}
\affiliation{State Key Laboratory of Quantum Optics Technologies and Devices, Institute of Laser Spectroscopy, Shanxi University, Taiyuan 030006, China}

\author{Jingxu Bai}
\affiliation{State Key Laboratory of Quantum Optics Technologies and Devices, Institute of Laser Spectroscopy, Shanxi University, Taiyuan 030006, China}
\affiliation{Collaborative Innovation Center of Extreme Optics, Shanxi University, Taiyuan 030006, China}

\author{Zhengyang Bai}
\email{zhybai@nju.edu.cn}
\affiliation{National Laboratory of Solid State Microstructures and School of Physics,
Collaborative Innovation Center of Advanced Microstructures, Nanjing University, Nanjing 210093, China} 

\author{Weibin Li}
\affiliation{School of Physics and Astronomy and Centre for the Mathematics and Theoretical Physics of Quantum Non-equilibrium Systems, University of Nottingham, Nottingham NG7 2RD, United Kingdom}

\author{Suotang Jia}
\affiliation{State Key Laboratory of Quantum Optics Technologies and Devices, Institute of Laser Spectroscopy, Shanxi University, Taiyuan 030006, China}
\affiliation{Collaborative Innovation Center of Extreme Optics, Shanxi University, Taiyuan 030006, China}

\author{Jianming Zhao}
\email{zhaojm@sxu.edu.cn}
\affiliation{State Key Laboratory of Quantum Optics Technologies and Devices, Institute of Laser Spectroscopy, Shanxi University, Taiyuan 030006, China}
\affiliation{Collaborative Innovation Center of Extreme Optics, Shanxi University, Taiyuan 030006, China}
	
\date{\today}

\begin{abstract}
Rydberg atoms have great potential for microwave (MW) measurements due to their high sensitivity, broad carrier bandwidth, and traceability. However, the narrow instantaneous bandwidth of the MW receiver limits its applications. Improving the instantaneous bandwidth of the receiver is an ongoing challenge. Here, we report on the achievement of an arbitrary instantaneous bandwidth MW receiver via a linear array of scalable Rydberg vapor cells with Stark comb, where the Stark comb consists of an MW
frequency comb (MFC) and a position-dependent Stark field. In the presence of the Stark field, the resonance MW transition frequency between two Rydberg states is position dependent, so that we can make each MFC line act as a local oscillator (LO) field to resonantly couple one Rydberg cell. Thus, each cell receives part of a broadband MW signal within its instantaneous bandwidth using atomic heterodyne detection, achieving the measurements of the broadband MW signal simultaneously. In our proof-of-principle experiment, we demonstrate the MW receiver with 210~MHz instantaneous bandwidth using an MFC field with 21 lines. Meanwhile, we achieve an overall sensitivity of 326.6~nVcm$^{-1}$Hz$^{-1/2}$. In principle, the method allows for achieving an arbitrary instantaneous
bandwidth of the receiver, provided we have enough MFC lines with enough power. Our work paves the way to design and develop a scalable MW receiver for applications in radar, communication, and spectrum monitoring.

\end{abstract}


\maketitle
\section{Introduction}  

The detection of microwave (MW) fields is of great significance in applications ranging from communication and electromagnetic compliance and safety, to defense and global navigation satellite system positioning~\cite{989947, Du_Swamy_2010, 9106073}. The application relies crucially on measuring broadband MW signals with high sensitivity~\cite{Teppati_Ferrero_Sayed_2013}. To date, MW field detection has been largely carried out with conventional electronic equipment. The inherent limitation of the solid state devices is that sensitivities are suffered to the thermal noise, and bandwidth is prone to the analog-digital converters~\cite{bolatkale2011,liu2022} and the resultant FFT processing burden. Recently, it has been shown that the latter achieves several gigahertz of instantaneous span.

Atom-based quantum sensors, on the other hand, have unique advantages of reproducibility, accuracy, and resolution. Inspired by this prospect, recent experiments have attempted to measure MW fields with a broad instantaneous bandwidth in an atomic vapor cell~\cite{Shi:2024ioe} and solid-state color centers~\cite{chipaux2015,carmiggelt2023}. In parallel, Rydberg atoms in atomic vapors have shown great potential in measuring MW fields. The advantage lies in their unique properties, such as large polarizability and MW transition dipole moment~\cite{gallagher1994rydberg}, as well as quantum coherence maintained by the optical electromagnetically induced transparency (EIT)~\cite{schlossberger2024a}. The abundant Rydberg states make it possible to achieve wide-frequency from DC to above THz~\cite{Fan2015,chen2022,li2023,lin2025}, including MW measurement~\cite{Sedlacek2012, Sedlacek13, SimonsMT2019, schmidt2025}, imaging~\cite{Holloway2014, Wade2017, downes2020b}, and wireless communication~\cite{Jiao2019, Song2019, Anderson2021, Holloway2019, Liu2022b}. Meanwhile, many methods have been developed to enhance the sensitivity. In the first milestone experiment based on the Rydberg-atom superheterodyne technique, the sensitivity reaches  55~nVcm$^{-1}$Hz$^{-1/2}$~\cite{Jing2020}. Later it has been improved to 30~nVcm$^{-1}$Hz$^{-1/2}$ by adding a repumping laser~\cite{Prajapati2021} and to 28.7~nVcm$^{-1}$Hz$^{-1/2}$ via the thermal resonance-enhanced transparency~\cite{hu2025}. Moreover, by placing the atom in a resonant MW cavity~\cite{Holloway:2022qwb,hao2022,sandidge2024}, the sensitivity is further improved down to 2.6~nVcm$^{-1}$Hz$^{-1/2}$~\cite{zhou2025a}.

Although important for detecting complex waveforms and receiving frequency spectra of a broadband MW signal, progress in the instantaneous bandwidth of Rydberg atom receivers is incremental. Various efforts, such as using lower probe power, higher coupling power, and narrow beam waist~\cite{hu2023, Yang:2023rvk, Shylla2025}, have achieved a bandwidth of $\pm$10.2~MHz~\cite{Yang:2023rvk} and can maintain the sensitivity~\cite{hu2023,manchaiah2025}. Utilizing an optical frequency comb, an instantaneous bandwidth of 12 $\pm$ 1~MHz is demonstrated~\cite{artusio-glimpse2024}. Recently, an instantaneous bandwidth of up to 54.6~MHz is realized with multi-dressed-state engineered Rydberg electrometry~\cite{yan2025}. 
Furthermore, a theoretical proposal based on the spatiotemporal multiplexed Rydberg MW receiver predicts the instantaneous bandwidth up to 100 MHz~\cite{knarr2023b}, although it has not been demonstrated experimentally.
The key reason is that long lifetimes in Rydberg states restrict refresh rates of ground-state atoms, which places limits on the instantaneous bandwidth~\cite{bohaichuk2022OriginsRydbergAtomElectrometer,2022TVandvideo}. The instantaneous bandwidth of Rydberg sensors is still limited to tens of MHz range, far below the GHz standard set by traditional sensors.

In this work, we present a scalable Rydberg vapor cell array designed to measure broadband MW signals. The system leverages functions of the Stark comb that consists of an MW frequency comb (MFC) and a position-dependent Stark field. The position-dependent Stark field is generated by a pair of electrodes, which create a gradient field tailored for spatially varying energy shifts. The MFC is engineered with 21 distinct frequency lines, spaced at 10 MHz intervals. This spacing corresponds to the instantaneous bandwidth of each Rydberg cell. Each MFC line acts as a local oscillator (LO) field to resonantly couple one local Rydberg cell, achieving detection of MW signal fields using atomic heterodyne detection. In our experiment, we change the relative position of a single cell
with respect to the electrodes along a linear direction, effectively replicating the functionality of a full Rydberg cell array.  We achieve an instantaneous bandwidth of up to 210~MHz and an overall sensitivity of 326.6~nVcm$^{-1}$Hz$^{-1/2}$ by making individual cell measurements. Furthermore, we validate its capability of reception by measuring a frequency-swept signal with a two-cell array. This approach enables scalable, high-sensitivity measurement of broadband MW signals while maintaining spatial and frequency control through the combined use of MFC and Stark fields.

\section {Principle of the Stark comb in a vapor cell array} 

\begin{figure}[htbp]
    \centering    
    \includegraphics[width=0.5\textwidth]{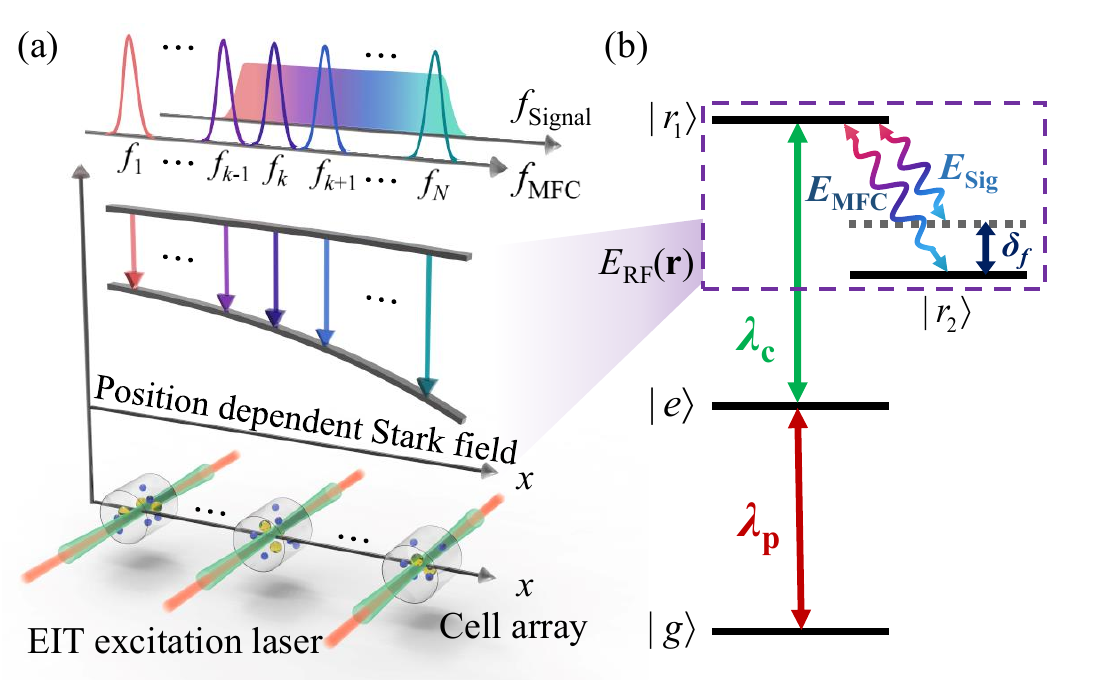}
    \caption{An arbitrary instantaneous bandwidth MW receiver. (a) Prototype of the receiver. The scalable Rydberg vapor cell array (bottom) is formed of $N$ individual cells along the $x$-axis. The Stark comb consists of an MFC and a position-dependent Stark field $E_{\text{RF}}(\bf r)$. (b) Relevant energy-level diagram for the MW receiver. The Rydberg state $|r_1\rangle$ is detected using EIT, where a probe laser ($\lambda_p$) and a coupling laser ($\lambda_c$) counterpropagate through the cell and drive the ground atoms $|g\rangle$ to the Rydberg state $|r_1\rangle$ via an intermediate state $|e\rangle$. The MFC and broadband frequency signal fields couple the Rydberg transition between $|r_1\rangle$ and $|r_2\rangle$. The resonance MW transition frequency between two Rydberg states is position-dependent, which is controlled by a position-dependent Stark field generated by an RF field.}
    \label{figure1}
\end{figure}

The mechanism of the arbitrary instantaneous bandwidth MW receiver is illustrated in Fig.~\ref{figure1}(a), where individual vapor cells are aligned along the $x$-axis.  In each cell, a probe and a coupling laser ($\lambda_p$ and $\lambda_c$) counter-propagate. These laser fields excite atoms from ground state $|g\rangle$ to Rydberg $|r_1\rangle$ via an intermediate state $|e\rangle$, forming an EIT configuration. The respective energy levels are depicted in Fig.~\ref{figure1}(b). 
The Stark comb comprises two key elements: an MFC and a position-dependent Stark field $E_{\text{RF}}(\bf r)$.
The MFC fields couple the state $|r_1\rangle$ with a different Rydberg state $|r_2\rangle$. The position-dependent Stark fields shift the Rydberg levels. As a result, the transition frequency, $\omega = \omega_0 + \frac{1}{2}(\alpha_1-\alpha_2)E_{\text{RF}}^2({\bf  r})$, between the Rydberg states depends on the cell position ${\bf r}$.  Here $\omega_0$ and $\alpha_j$ ($j=1,2$) are the field-free transition frequency and polarizability of state $|r_j\rangle$.  By moving the cell,  the transition frequency can be adjusted to match one of the MFC lines. In other words, this MFC line serves as an LO field that resonantly couples the respective Rydberg cell, as illustrated in the middle of Fig.~\ref{figure1}(a). Other MFC lines, as non-resonant fields, couple two Rydberg states. When a broadband MW input signal field is applied, shown as the top of Fig.~\ref{figure1}(a), the input signal and resonant LO fields are coupled through the Rydberg state transitions~\cite{Jing2020, SimonsMT2019} within its instantaneous bandwidth, resulting in beat notes in individual cells. The beat notes are read out from the transmission spectrum. The spatially dependent atomic response of this array effectively acts as a receiver for a broadband MW field in parallel. The response of other beat notes generated by the non-resonant MFC lines is weak and negligible.  

\begin{figure*}[htbp]
    \centering    \includegraphics[width=1\textwidth]{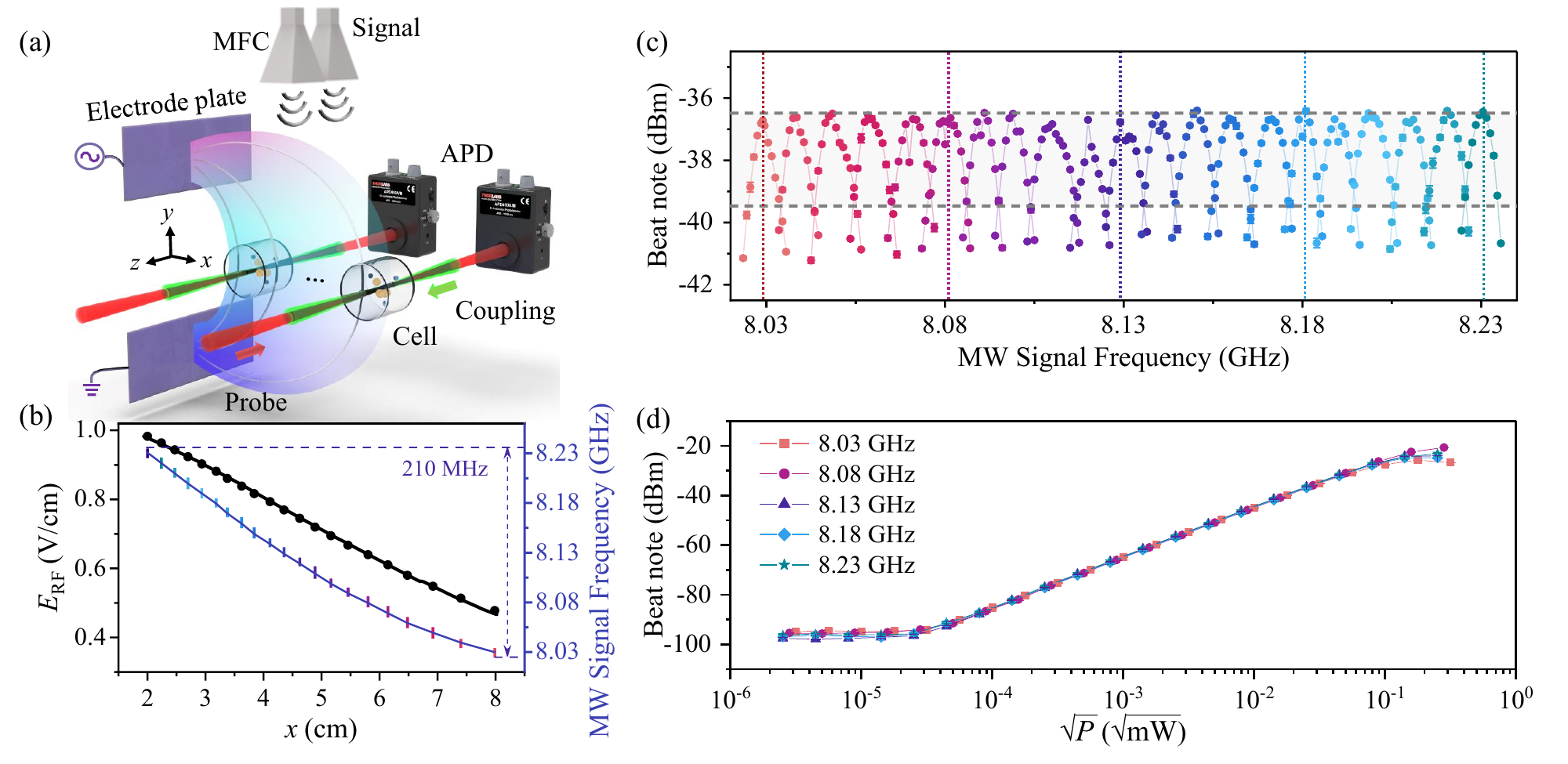}
    \caption{(a) Schematic of the experimental setup. An RF electric field is applied to a pair of aluminum electrode plates to generate the position-dependent Stark fields. An MFC as LO fields and a single-tone MW signal are fed to two identical horn antennas, respectively, and simultaneously incident on the Rydberg MW receiver. A cell is placed on the $x$-axis, and a 509~nm Rydberg coupling laser and a 852~nm probe laser counterpropagate through the cell and drive the ground state to the Rydberg state, forming the Rydberg EIT. The transmission of the probe laser is detected by an APD. The electrodes are moved to change the relative position between the cell and the electrodes along the $x$-axis. (b) The black solid line and dots show the calculated and experimental data of the RF electric field strength along the x-axis. The \textit{x}-coordinate of the black dots presents the chosen 21 positions of cells in the experiment. The closet minimum spacing between two cells is 0.23~cm. The blue solid line presents the calculated resonant MW transition frequency between two Rydberg states along the $x$-axis. Each colored vertical line point demonstrates the range of instantaneous bandwidth at the corresponding \textit{x} position, which is extracted from (c), and their colors correspond to the colors in (c). The total instantaneous bandwidth is 210~MHz, marked by the dashed lines, i.e., enabling measurement of a 210~MHz broadband MW signal simultaneously.
    (c) Instantaneous bandwidth of one cell at 21 different positions. The 3~dB bandwidth is labeled by the horizontal dashed grey lines. We measure the signal frequency with a 210~MHz bandwidth from 8.025-8.235~GHz. (d) Power of the beat note as a function of the signal field strength at the indicated signal frequency. The receiver has the same response at the same input power in the linear dynamic range. The flat region indicates that the beat note power reaches the noise level.}
    \label{figure2}
\end{figure*}

The parallel measurement of multiple MFC lines is demonstrated using the example of two Rydberg vapor cells. 
To provide a clear illustration of the working principle, we propose a gradient Stark field to
shift the MW transition frequency, making it a function of the cell position $x$, i.e., $\frac{1}{2}(\alpha_1-\alpha_2)E_{\text{RF}}^2(x)$.
We assume that each cell has an instantaneous bandwidth of $\pm f_{\rm IBW}$, as commonly used in current experiments. Accordingly, the frequency interval of the MFC, $\Delta f_{\rm MFC}$, is around 2$f_{\rm IBW}$, i.e, $\Delta f_{\rm MFC}\sim 2f_{\rm IBW}$. This ensures that the instantaneous bandwidth connection between each cell is continuous.
For achieving high efficiency receiving, we set $\Delta f_{\rm MFC}= 2f_{\rm IBW}$. Geometrically, the first and the second cell, located at positions $x_1$ and $x_2$, are resonantly coupled to the first and second MFC lines, with frequencies $f_1$ and $f_2$ ($f_2>f_1$), respectively. 
This configuration yields an enlarged bandwidth of $4f_{\rm IBW}$ when $f_2-f_1 = 2f_{\rm IBW}$, achieved by precisely tuning the positions of the two cells to establish the required frequency interval.

The simple scenario showcases its scalability potential, enabling extension to larger array configurations. To do so, we place an array of cells along the Stark field, where each cell resonantly couples to one MFC line ($f_k$) sequentially. If we have an array with $N$ cells and $N$ MFC lines, this allows to detect MW field signals with a bandwidth  $N\times 2f_{\rm IBW}$ simultaneously. The expansion of the bandwidth increases linearly with $N$ in this arrangement. In principle, we can achieve an arbitrary instantaneous bandwidth MW receiver using the scalable Rydberg vapor cell array with Stark comb, provided that we design properly the position-dependent Stark fields, and have enough MFC lines with abundant power. Moreover, utilizing the wafer-level fabrication method~\cite{artusio-glimpse2025}, we can achieve a miniaturized MW receiver with an integrated vapor cell array.

\section {Experimental realization} 

In our experiment, we realize a proof-of-principle demonstration by changing the relative position of a single cell with respect to the electrodes along the \textit{x}-axis in the spatially dependent Stark field.  The sketch of the experimental setup is illustrated in Fig.~\ref{figure2}(a). The cell has a size of $\phi$ 2.5~cm $\times$ 2~cm, and is placed along the $x$-axis. A 509~nm Rydberg coupling laser (Rabi frequency $\Omega_c = 2\pi \times$ 16.1~MHz ) 
and a 852~nm probe laser (Rabi frequency $\Omega_p = 2\pi \times$ 6.9~MHz $\sim$ 9.3~MHz) counterpropagate through the cell, driving the ground state $|g\rangle = |6S_{1/2}, F = 4\rangle$ to the Rydberg state $|r_1\rangle = |45D_{5/2}\rangle$ via an intermediate state $|e\rangle = |6P_{3/2}, F^\prime = 5\rangle$. The probe and coupling lasers keep co-linear polarization along the $x$-axis, and their $1/e^2$ beam waists $\omega_p$ and $\omega_c$ are 70~$\mu$m and 90~$\mu$m, respectively. 

To build the Stark comb, we generate the position-dependent Stark field by applying a 120~MHz RF field to a pair of aluminum electrode plates (size of 12\,\text{cm} $\times$ 7.5\,\text{cm} $\times$ 0.1\,\text{cm} with a separation of 12\,\text{cm} between two edges) that are placed symmetrically on the \textit{z}-axis in the $yz$ plane, so that the strength of the Stark field varies along $x$-axis. The field strength is shown by the black curve (simulation) and dots (experimental measurements) in Fig.~\ref{figure2}(b). Our MFC outputs 21 frequency lines with an interval of 10~MHz between two neighboring lines at a center frequency of 8.13~GHz, which drives the transition $|r_1\rangle\to|r_2\rangle = |46P_{3/2}\rangle$. The MW resonant transition frequency between $|r_1\rangle$ and $|r_2\rangle$ without Stark fields is 7.97~GHz. We set the interval to be 10~MHz as the instantaneous bandwidth of a single cell is $\pm$~5MHz.  Numbers of MFC lines are mainly limited by the power of the generator. 

The MFC as LO fields and a single-tone MW signal field are fed to two identical horn antennas, respectively. They simultaneously incident on the receiver with both polarizations the same as the laser beams. The signal field is mixed with one of the MFC lines that is closest to its frequency with a frequency difference of $\delta_f$ via Rydberg atoms, resulting in a beat note that is detected by avalanche photodetectors (APD) after each Rydberg cell~\cite{Jing2020, SimonsM2019, gordon2019}.
In the experiment, we change the relative position of the cell with respect to the electrodes along the $x$-axis. This allows for changing the MW transition frequency of 10~MHz at the next cell location.

In Fig.~\ref{figure2}(b), the \textit{x}-coordinate (black dots) gives the relative position between the cell and electrodes in the experiment. At each matched position, we record the data, then we stitch all the data together, which is equivalent to placing multiple cells sequentially. The blue curve in Fig.~\ref{figure2}(b) presents the calculated resonant MW transition frequency between the two Rydberg states along the \textit{x} axis.
Note that the Stark fields cause the $m_j$-dependent Stark shifts and splittings of $|45D_{5/2}\rangle$ and $|46P_{3/2}\rangle$ states~\cite{miller2016,jiao2017}. We choose states of $|45D_{5/2}, m_j=5/2\rangle$ and $|46P_{3/2}, m_j=3/2\rangle$ to couple with the MFC field. This is because the selected $|45D_{5/2}, m_j=5/2\rangle$ has the maximum EIT spectrum when the polarization of the Stark field (\textit{z}-axis) is perpendicular to that of the excitation lasers (\textit{x}-axis)~\cite{li2023}. Meanwhile, at each matched position, the frequency of the coupling laser is tuned to compensate for the Stark shift of the $|45D_{5/2}, m_j=5/2\rangle$ state. 

\section {Results} 

The instantaneous bandwidth of the receiver is defined as the frequency range corresponding
to a 3~dB power attenuation of the output beat notes. First, we measure the instantaneous bandwidth of an individual cell at each chosen position by varying the signal MW frequency, while fixing the total output power of MFC at 11~dBm and the MW signal field at -30~dBm. The power of the output beat note is analyzed by a spectrum analyzer. We record the data within the span bandwidth of 5~MHz in the spectrum analyzer, so that the beat notes are from the beating between the signal field and its closest MFC field line. As shown in Fig.~\ref{figure2}(c), data points with the same colors present the beat note power taken from the cell at a given position of the Stark field. There are 21 different sets of colors, corresponding to 21 positions [the \textit{x}-coordinate of the
black dots in Fig.~\ref{figure2}(b)]. From left to right, the Rydberg level detuning is increased. The peak point of each data set is taken with $\delta_f=$ 500~kHz between the MW signal field and its closest MFC line. The 3~dB line is labeled by the horizontal dashed grey lines. Note that the absolute frequency of the signal can be obtained using the method demonstrated in Ref.~\cite{zhang2022}.

The profile of each data set clearly shows that the power of the beat note decreases rapidly when the frequency of the signal field is away from its closest MFC comb line, while after its 3~dB bandwidth, the beat note becomes higher again at the next position. After we stitch all the data together, we can achieve the total instantaneous bandwidth of 210~MHz within a 3~dB bandwidth. In addition, we extract the instantaneous bandwidth at each position and plot them as vertical line points in Fig.~\ref{figure2}(b). The results show that the centers of all the vertical line points are almost located on the theoretical line, representing the resonant MW transition frequencies, and the range of each line point corresponds to its instantaneous bandwidth. 

In our measurement, each peak of beat note power in Fig.~\ref{figure2}(c) has almost identical -36.5~dBm. This is achieved by optimizing the Rabi frequency of the probe laser in the range of $\Omega_p = 2\pi \times$ 6.9~MHz $\sim$ 9.3~MHz and the strength of each MFC line. As a result, the system responds homogeneously for different frequency signal fields with the same power. To further verify this, we measure the response of the system as a function of the signal field strength at each position. Fig.~\ref{figure2}(d) demonstrates the power of the beat note as a function of the signal field strength at the indicated signal frequency [marked by the vertical dotted line in (c)]. In each measurement, the signal frequency has a 500~kHz detuning relative to its closest MFC line. The results show that the response of the system overlaps at the same input signal power in the linear dynamic range.

\begin{figure}[htbp]
    \centering      \includegraphics[width=0.5\textwidth]{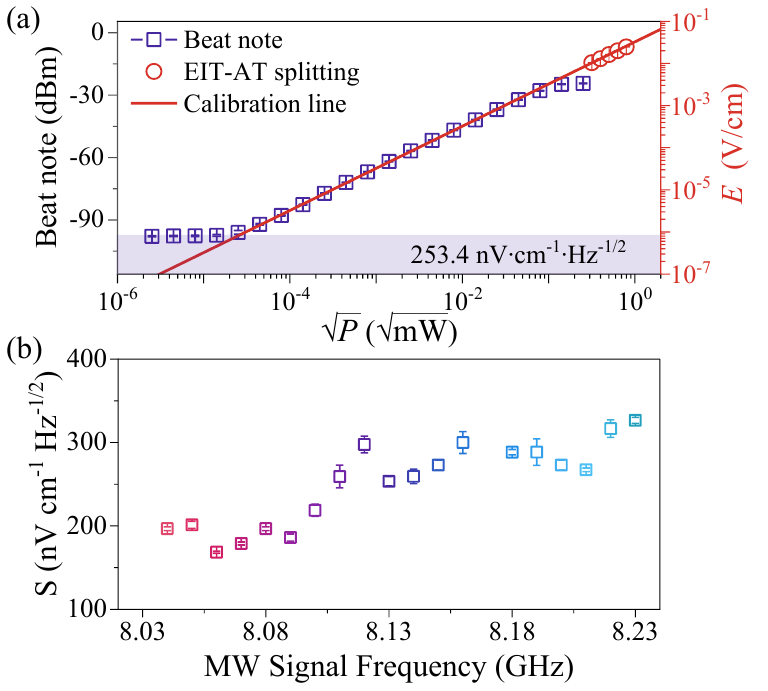}
    \caption{Sensitivity for the cell with MW resonant frequency at 8.13~GHz. The purple region represents the noise level. The red circles show the EIT-AT splitting in a strong field region, and the red solid line shows the calibrated electric field. (b) Sensitivity at different positions. Each colored point corresponds to the color in Fig.~\ref{figure2}(c).}
    \label{figure3}
\end{figure}

Subsequently, we measure the sensitivity of the MW receiver. In Fig.~\ref{figure3}(a), we show an example where the cell is located at the position with transition frequency 8.13~GHz, corresponding to the center line of MFC as the LO field. The signal frequency has a 500~kHz detuning relative to the LO field. The purple squares present the power of the beat notes as a function of the signal field strength. The EIT-AT splitting in a strong field region (red circles) and the far-field formula $E_{\mathrm{FF}} = {F\sqrt{30P G}}/{d}$ (red solid line) are used to determine the strength of the signal fields, where $G$ is a gain factor of the antenna, $d$ is the distance between the horn antenna port and the center of the cesium cell, and $F$ is the cell perturbation factor. A minimum detectable electric field $E_{\mathrm{det}}=$ 798.2~nV/cm is determined when the power of the beat note reaches the noise level, and the sensitivity is then obtained via $S = E_{\mathrm{det}} \cdot \sqrt{T}$, where $T = $ 0.1~s  is the measurement time. In this case, the sensitivity is found to be 253.4~nVcm$^{-1}$Hz$^{-1/2}$. 

Similar to the previous example,  we perform a series of measurements and obtain the sensitivity of the Rydberg cell at different positions, depicted in Fig.~\ref{figure3}(b). The results show that the sensitivity gains a trend of degradation with the increasing Rydberg level tuning. This is due to the fact that when increasing the Rydberg level tuning, the MW transition dipole moment decreases. From the data, we identify the worst sensitivity to be 326.6~nVcm$^{-1}$Hz$^{-1/2}$, which gives the overall sensitivity of the receiver.

\begin{figure}[htbp]
    \centering
    \includegraphics[width=0.5\textwidth]{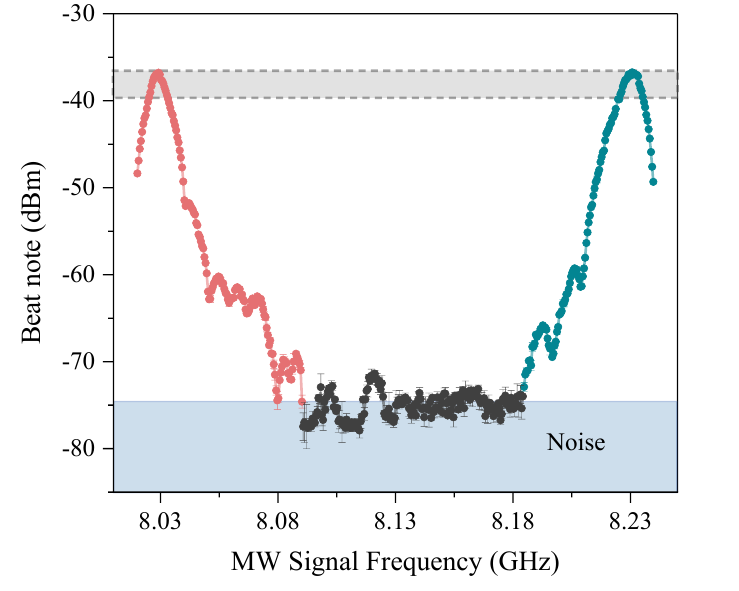}
    \caption{Beat note signal for measuring two cells simultaneously. The cells are placed on the most left (red dots) and most right (green dots) ends of the Stark field. 
     The black dots show the noise level. The 3~dB bandwidth is labeled by the horizontal dashed lines.}
    \label{figure4}
\end{figure}

Finally, we carry out a dual-path reception experiment by placing two cells on the most left ($x=$ 2~cm) and the most right ($x=$ 7.98~cm). This setting corresponds to an equivalent example of scalability. The MW frequencies are resonant with the last (8.23~GHz) and first (8.03~GHz) MFC lines, respectively. Due to the different Stark shifts in the two positions, the two coupling lasers have a frequency shift of 13.4~MHz. After locking the two coupling lasers to their EIT peak, we perform the reception of a frequency-swept signal to demonstrate its capabilities. In Fig.~\ref{figure4}, we demonstrate the measured beat notes by sweeping the signal MW frequency from 8.02-8.24~GHz. The results show the frequency difference between the two peaks is 200~MHz, which validates the equivalent instantaneous bandwidth of 210~MHz by adding to its own 3~dB bandwidth. The agreement shows the capability of receiving MW field with a large instantaneous bandwidth. This is the first time, as far as we can tell, to realize the highest instantaneous bandwidth measurement with the vapor cell setting. It moreover shows the potential to achieve even higher bandwidth with a scalable array of vapor cells. Based on the demonstrated principle, we will set up 21 integrated cells to achieve 210~MHz broadband MW signal reception simultaneously and continuously in the near term.

\section{Conclusion and discussion}
We have demonstrated an MW receiver for the measurement of broadband MW signals via the scalable Rydberg vapor cell array with Stark comb. Our system achieves an instantaneous bandwidth of 210~MHz and an overall sensitivity of 326.6~nVcm$^{-1}$Hz$^{-1/2}$. The capabilities are demonstrated by measuring a frequency-swept MW signal. The sensitivity of our MW receiver can be improved by using higher Rydberg states~\cite{cai2023}. The instantaneous bandwidth, on the other hand, can be enlarged by increasing the number of MFC lines and cells, together with the appropriate position-dependent Stark field. When combining the multi-dressed state method in individual cells~\citep{yan2025}, our approach is able to achieve an instantaneous bandwidth of $>$ 1~GHz without major changes in the setup. To achieve a larger instantaneous bandwidth, the MW transition frequency between two Rydberg states should be large enough. The required Stark field may cause state mixing and shift the energy spacing out of resonance in nearby cells. This can be mitigated by employing different pairs of Rydberg states in neighboring cells, since the laser excitation in the $N$ vapor cells is mutually independent.  This allows the receiver to connect the MW signal frequency continuously. Our study provides a scalable, economic approach to achieve an arbitrary instantaneous
bandwidth of the Rydberg atom-based MW receiver. 
This enables broadband signal detection, frequency-hopping communications, and real-time MW signal monitoring — capabilities that are critical for applications such as radar, 5G millimeter-wave communications, electronic reconnaissance, satellite communications, and cosmological research.

\section{Acknowledgment} 
The work is supported by the National Natural Science Foundation of China (No. U2341211, No. 62175136, No. 12241408, No. 12120101004, No. 12274131); Innovation Program for Quantum Science and Technology (No. 2023ZD0300902);  Fundamental Research Program of Shanxi Province (No. 202303021224007); and the 1331 project of Shanxi Province. W.L. acknowledges financial support from the EPSRC (No. EP/W015641/1) and the Going Global Partnerships Programme of the British Council (No. IND/CONT/G/22-23/26).

\bibliography{main}

\end{document}